\documentclass[aps,prl,twocolumn,showpacs]{revtex4}
\usepackage{graphicx}\usepackage{subfigure}
\newcommand{\beq}{\begin{eqnarray}}
\newcommand{\eeq}{\end{eqnarray}}
\newcommand{\ket}[1]{\vert  #1 \rangle }
\newcommand{\bra}[1]{\langle #1 \vert  }
\newcommand{\Tr}[1]{  {\rm Tr} \left\{ #1 \right\} }
\begin{document}

\preprint{1}
\title{Witnessing entanglement beyond decoherence}
\author {Felipe Caycedo-Soler, Ferney J. Rodr\'{i}guez and Luis Quiroga}
\affiliation{Departamento de F\'{i}sica, Universidad de los Andes,
A.A. 4976 Bogot\'a, D.C.,Colombia}
\date{\today}
\begin{abstract}
 We address the witnessing of quantum correlations beyond the limits imposed by an ensemble statistical average.  By relying upon the continuous observation of
 a single quantum open system under the action of classical or quantum noise, we show how the statistical distribution of outcomes of physically measurable quantities provides enough information to detect entanglement in situations where a simple averaging  procedure yields to a vanishing value. In this way, we are able to specify
 the meaning of decoherence when single read-out measurements are available, to demonstrate that these distributions are capable to predict/detect  entanglement well beyond the  ensemble average provided by a dephasing-sensitive mixed state.
 Few-spin and fluorescent systems are used to illustrate entanglement survival when a mixed state average predicts none or
 under the conditions where the sudden-death of entanglement is expected.
\end{abstract}
\maketitle

{\it Introduction.-} The quantum mechanics most highlighted feature is non-locality, the cornerstone of  entanglement, called by Schr\"odinger ``the characteristic trait of quantum mechanics" \cite{schroedinger}. Locality emerges when a system interacts with an environment (open system),   the former usually  described  by a statistical mean, i.e.  the system's reduced mixed state $\rho$,  supported by disregarding beyond the average, the effect from numerous degrees of freedom of the latter. As an average, information regarding the spectrum of possible states (decompositions or unravelings of $\rho$) is degraded, a fact  that makes  mixed state  entanglement quantification  more complicated and less well understood than that of pure states \cite{wootters}.  Recently, a better understanding of quantum correlations in a mixed state has become feasible by addressing the entanglement directly linked to physically realizable decompositions \cite{carmichael, viviescas2}.  Additionally, the increasing interest to identify measurable observables able to experimentally witness entanglement  (entanglement witnesses, EWs) \cite{plenio,vedral}, jointly with an increased resolution to monitor the dynamics of single quantum open systems \cite{experiments,almeida},  make it possible to detect quantum non-locality features, beyond the  smeared out average environment fluctuations,  in optical,  atomic/molecular and condensed matter physics.

In this Letter  we address the crucial issue of detecting quantum correlations through the  statistical distribution of EW outcomes captured from continuous environment observation of open quantum systems.  Firstly,  the possibility to witness quantum correlations with dephasing arising from classical noise is studied beyond ensemble averages at the level of realizations from a single quantum system, to show that in general the  dephasing mechanisms do not always suppress quantum correlations,  but even more surprisingly, may escalate them up.  Secondly, an extension to  fully quantum entangled system-environment situations is  considered to show how  witnessing of quantum correlations, beyond the mixed-state  entanglement-sudden-death (ESD) \cite{eberly,almeida} framework, strengthens the relevance  of a proper description of open quantum systems where single read-out measurements are becoming commonplace \cite{PRA,lukin}.

The mixed state $\rho=\sum_k p_k\vert \psi_k\rangle\langle \psi_k\vert$,   is appropriate whenever no knowledge on which  state $\vert \psi_k\rangle$ the system is really in, with infinite decompositions or sets  $\{p_k,\vert \psi_k\rangle\}$, each yielding to a  different entanglement mean value. A commonly used measure of entanglement for bipartite systems in mixed states, the concurrence $C=\mbox{Max[}0,\lambda_1-\sum\lambda_i]$, where $\lambda_1>\lambda_i$ are square root  eigenvalues of $\rho\tilde{\rho}$, $\tilde{\rho}=\sigma_y\otimes\sigma_y\rho^*\sigma_y\otimes\sigma_y$  ($\sigma$ are Pauli spin matrices throughout this paper), addresses the asymptotic ratio $m/n$ of minimal resource of $m$ ebits required to form an ensemble of $n\rightarrow\infty$ pairs described by  $\rho$, and singles out the minimally entangled decomposition of $\rho$, $C=\mbox{min}\langle C\rangle$ \cite{wootters,viviescas2}.
For instance,  an ensemble of pairs of qubits (QBs) with  basis states $\ket{\uparrow},\ket{\downarrow}$ prepared  in (dephased) fully entangled states $(\ket{\uparrow\downarrow}+e^{i\phi}\ket{\downarrow\uparrow})/\sqrt{2}$ with  $\phi$ uniformly distributed  over (0,2$\pi$], has an average entanglement $\langle C\rangle=1$ that sharply contrasts with the result $C=0$ from the purely classical mixed-state  $\rho=\frac{1}{2}(\ket{\uparrow\downarrow}\bra{\uparrow\downarrow}+\ket{\downarrow\uparrow}\bra{\downarrow\uparrow})$, which arises from  averaging this ensemble of dephased states.    Moreover,  any ensemble averaged physical quantity (therefore, any EW)  is, as a matter of fact, invariant to any decomposition. Consequently, it will not unveil the differences in entanglement with, say,  a half  $\ket{\uparrow\downarrow}$-half $\ket{\downarrow\uparrow}$ ensemble  ($\langle C\rangle=0$) that yields to the same fully mixed state. In order to illustrate the relevance of properly relying in EWs statistical distributions with realistic physical unravelings from which ensemble averages are made upon, spin as well as fluorescent systems will be addressed below.

\begin{figure}
\advance\rightskip-1cm
\includegraphics[width= 2.6 cm]{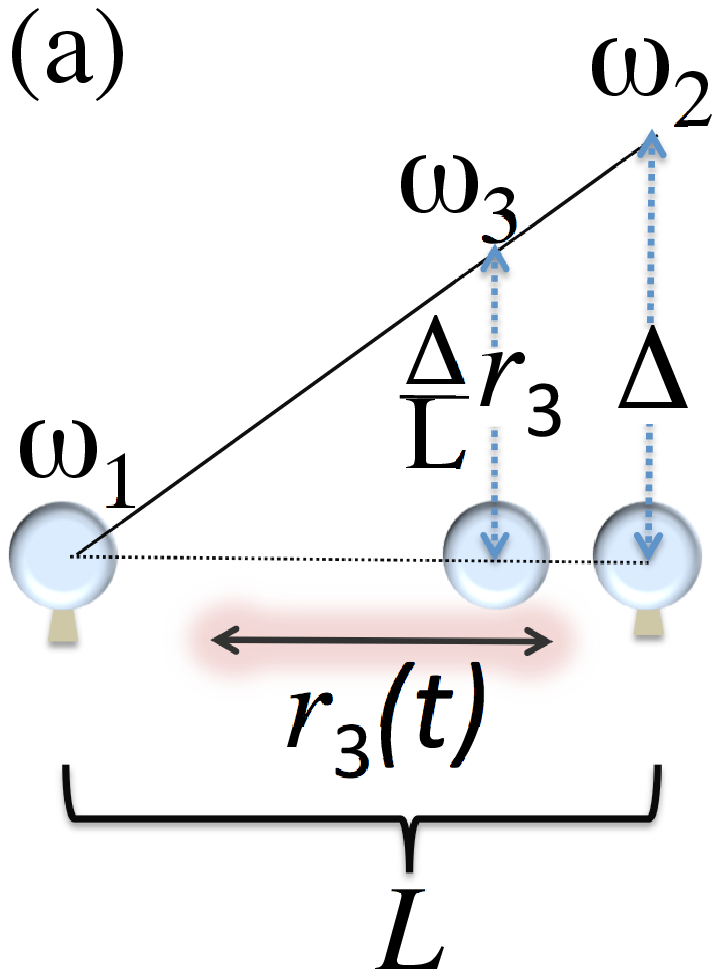}
\hspace{-.1 cm}
\includegraphics[width= 5.6 cm]{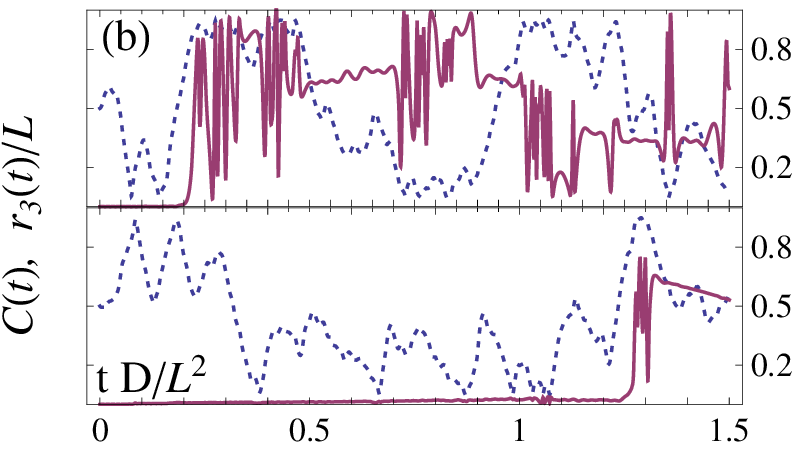}\\
\advance\rightskip1cm
\includegraphics[width= 4 cm]{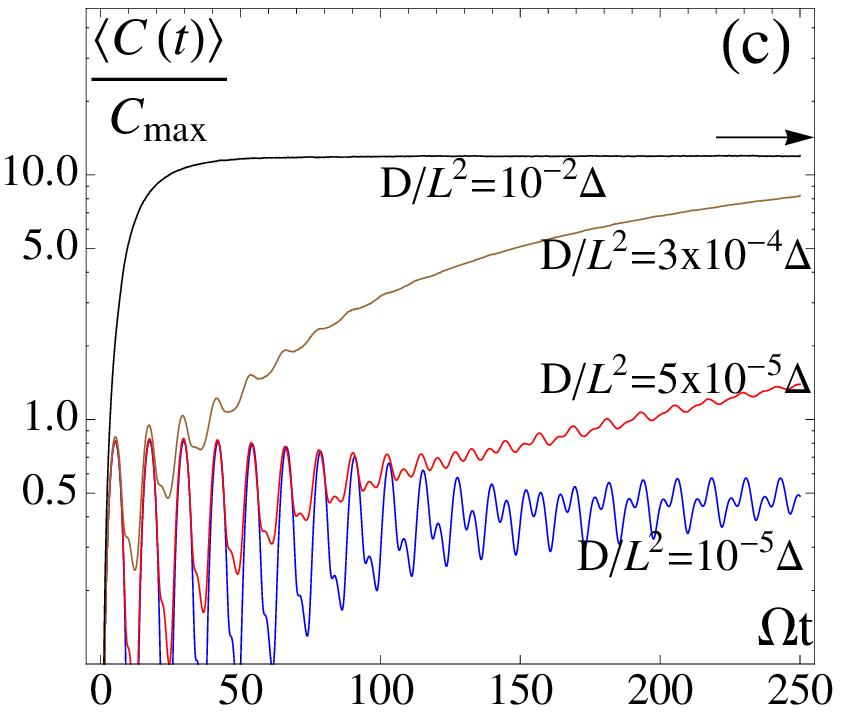}
\includegraphics[width= 4 cm]{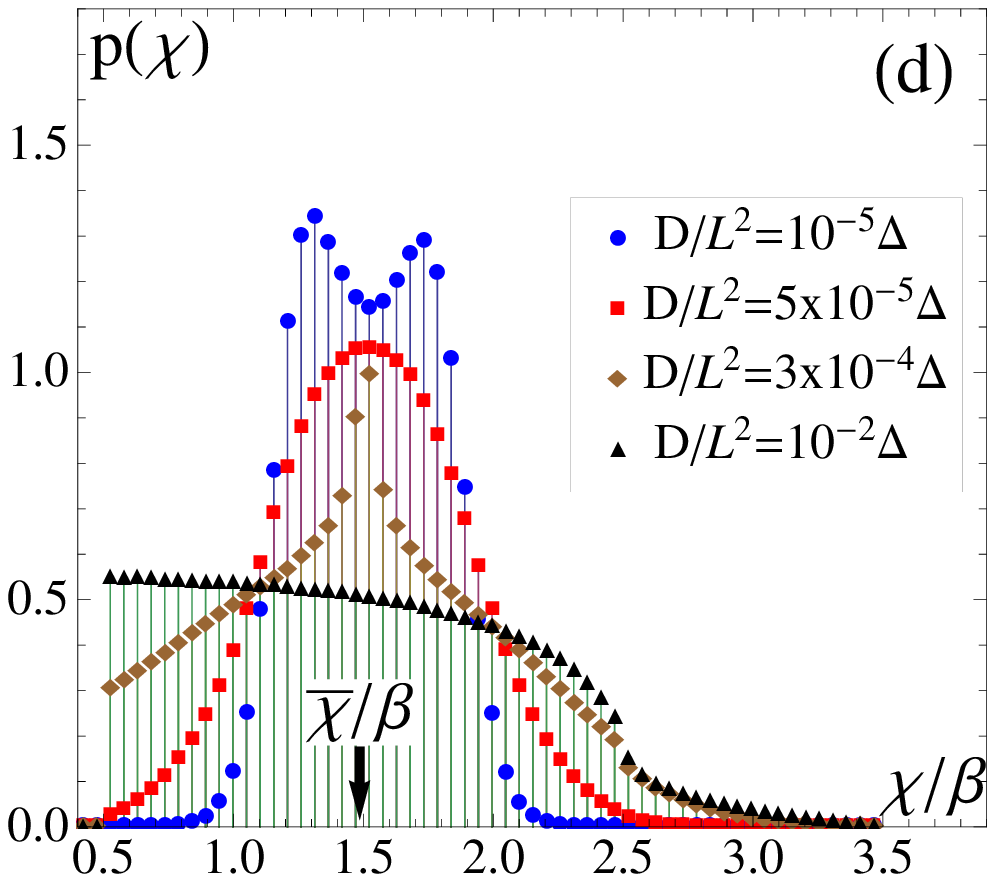}
\caption{(Color online)  (a)  The classical noise physical set up: two static particles interacting via dipolar coupling with a diffusing third particle (see text). (b) Two typical trajectories $r_3(t)$ (dots) with the corresponding realizations for concurrence $C$ (continuous)  between spins 1+2.
The enhancement of quantum correlations: the effect of diffusivity $D$ is shown  in (c),  $\langle C\rangle/C_{\mbox{\footnotesize{max}}}$ reaching closely the average concurrence value for resonant interaction $C_{\mbox{\footnotesize{mean}}}$ (arrow); in (d) the susceptibility distribution $\chi/\beta$ is shown where the  value $\bar{\chi}/\beta=1.494\pm0.002$ predicted by the average state $\rho$ for $D/L^2=\Delta$ is highlighted ($\Omega t=250$). In all plots $ \Delta=100g_0$, $\ket{\psi(0)}=\ket{\uparrow_1\downarrow_2\downarrow_3}$. In (c) and (d) $5\times10^4$ realizations have been used.}\label{brownian}
\end{figure}

{\it Classical noise.-}  Consider a system of $N$ interacting $1/2$-spins described by a Hamiltonian ($\hbar=1$) such as
 \beq
H_s=\sum_i^N\frac{\omega_i(r_i)}{2}\sigma^z_i+\sum_{i\neq j}^N \frac{g(r_{i,j})}{2}\sigma^+_i\sigma^-_j,\label{Hs}
\eeq
with distance dependent couplings,  specifically dipole-like interactions $g(r_{i,j})=g_0/(\frac{\vert \vec{r}_{i}-\vec{r}_j\vert}{L})^3$ subject to classical noise coming from  the random dynamics of the spins relative positions  $r_{i,j}=\vert \vec {r}_i-\vec{r}_j\vert$. First, in the absence of noise or static case, for $N=2$  ($r_{1,2}=L$) and initial state $\ket{\uparrow_1\downarrow_2}$ (subindex labels each spin), in resonance $\omega_2-\omega_1=\Delta=0$, the concurrence is found to oscillate with a frequency $g_0$ with a full cycle average $C_{\mbox{\footnotesize{mean}}}=2/\pi$. On the other hand, far from resonance
i.e. $\Delta\gg g_0$, the concurrence oscillates with frequency $\Omega=\sqrt{\Delta^2+g_0^2}\approx \Delta$ reaching a fairly small maximum value of  $C_{\mbox{\footnotesize{max}}}=2 g_0 \Delta/\Omega^2\approx2g_0/\Delta$. 
 Classical noise is now included allowing a third particle (3) to elastically bounce  between the static particles 1 and 2, such that  $r_1 <\langle r_3\rangle<r_2$, while it diffuses following a Brownian motion in  an inhomogeneous external magnetic field, i.e. $\omega_3(r_3)\sim \langle r_3\rangle\Delta/L $ (Fig\ref{brownian}(a)). If the translational wave packet extension of the mobile particle can be disregarded,  its diffusion is described by a Langevin equation  ($\langle r_3\rangle=r_3$): $\ddot{r_3}=-\frac{1}{\tau} \dot{ r_3}+\sqrt{w}F(t)$, where $F(t)$ is an unit variance  white Gaussian noise, $\tau$ and $w$ are respectively, relaxation time and diffusion constant, related to the diffusivity by $D=w\tau^2/2$  \cite{gillespie}.  The noisy reservoir,  here represented by the Brownian particle, leads to dephasing among  pure state realizations.  The concurrence between static spins 1 and 2 is calculated relying upon exact numerical simulations for  Brownian motion  \cite{gillespie} to find the position of particle 3 on a coarse-grained time-scale inherent to nanometric particles  ($\tau\ll1/\Omega$). This last fact implies a single translational time-scale $L^2/D$ \cite{thesisfcs} able to wrap several visits of particle 3 to the static ones,  useful to enhance  entanglement between particles 1 and 2 when they are out of resonance (Fig.\ref{brownian}(b)). Figure \ref{brownian}(c) shows  that  the noisy  motion of particle 3 improves both resonance and interaction strength to closely reach  the noiseless resonant situation and remarkably to enhance over one order of magnitude in {\it average}, the {\it maximum} fixed particles original off-resonant entanglement. The magnetic  susceptibility $\chi$ is an EW,  whenever the inequality $\chi/\beta<N s$ \cite{vedral} ($\beta$ is the inverse of temperature times Boltzmann constant, $s=\frac{1}{2}$ for QBs) holds.  A saturation value of $\chi/\beta=\frac{3}{2}$ is found for any  classical mixture of states   $\ket{\psi}=\alpha\ket{\uparrow_1\downarrow_2\downarrow_3}+\gamma\ket{\downarrow_1\uparrow_2\downarrow_3}+\delta\ket{\downarrow_1\downarrow_2\uparrow_3}$. The closeness of the ensemble average $\bar{\chi}/\beta$  to $\frac{3}{2}$,  is emphasized  in Fig.\ref{brownian}(d), to reflect  the fact that the simple statistical average (from nearly a classical mixture) is, in practice, unable to distinguish  the enhancement of quantum correlations when noise is present. The maximum $\chi/\beta=\frac{7}{2}$ happens when  $\alpha=\gamma=\delta$ , and bounds a domain  down to  $\chi/\beta=\frac{3}{2}$ shared with more general three QBs classical mixtures. However, the  distribution of susceptibility from realizations  shows (Fig.\ref{brownian}(d)) that the  main effect of noise is to enlarge the tail of the distribution on the generally classical forbidden region down to  $\chi/\beta=\frac{1}{2}$  (arising from all permutations $\{\alpha,\gamma,\delta\}$ such that $\alpha=\gamma=-\frac{1}{2}\delta$). The fact that no conclusion on quantum correlations is possible from the EW  ensemble statistical average   (susceptibility) due to the mixed state  dephasing-entanglement trade off, is relieved when access to the EW statistical distribution is attained, by accounting on the additional information available from the detailed observation of the environment-induced fluctuations.

 {\it Quantum noise.-}  The witnessing of entanglement to overcome the inevitable dephasing rooted in ensemble averages, can be extended to a fully entangled system-environment state $\ket{\Psi}$, which  traced out (hence averaged out) over the bath, provides the reduced density operator of a mixed state, $\rho=\Tr{\ket{\Psi}\bra{\Psi}}_{\mbox{\footnotesize{env}}}$. The Schr\"odinger-HJW theorem  \cite{HJW}, states that (infinite) sets of reservoir measure-operators  $\{F_k^{1/2}\}$ exist without disturbing the average state of the system, and whose outcomes make visible a specific decomposition of the mixed state $p_k\ket{\psi_k}\langle \psi_k\vert= \Tr{\mathbf{I}_{\mbox{\footnotesize{sys}}}\otimes F_k^{1/2}\ket{\Psi}\langle \Psi\vert F_k^{1/2}\otimes\mathbf{I}_{\mbox{\footnotesize{sys}}}}_{\mbox{\footnotesize{env}}}$.  Consequently, this last result allows to project onto the  pure state $\ket{\psi_k}$ in which the system is really in, based upon the record of environment dynamics. The measure-operators produce a change in the system's state when the system and environment are entangled, as resolved by the so-called Kraus operators $\Omega_k(t)$, which fulfill $\rho(t)=\sum_k \Omega^\dagger_k(t)\rho(0)\Omega_k(t)$ and $\sum_k \Omega^\dagger_k(t)\Omega_k(t)=\mathbf{I}$.  The Kraus operators allow to describe exclusive evolutions $\ket{\psi(t+\Delta t)}_k=\Omega_k(\Delta t)\ket{\psi(t)}$- preserving purity  of states from unit efficiency environment measures \cite{wiseman}- conditioned on a particular outcome $k$ arising with probability $p_k=\langle\Omega_k(\Delta t)^\dagger\Omega_k(\Delta t)\rangle_{\psi(t)}$. This evolution (formally $\Delta t$ is greater than the bath correlation time)  proceeds with no memory under the Born-Markov approximation, according to a Lindblad  type master equation
 \begin{equation}\label{meq}
\partial_t\rho=i[H_0,\rho]+\sum_i L_i\rho L_i^\dagger -\frac{1}{2}\{L_i^\dagger L_i,\rho\},
\end{equation}
where the operators $L_{i}$ are associated to the irreversible effect of the environment on the system. Physically, spontaneous emission arising from the interaction of the quantized electromagnetic vacuum with an excited atom or QB fulfills the Born-Markov approximation. The ensemble averaged relaxation from the excited state $\ket{\uparrow}$ to the ground state $\ket{\downarrow}$ of a QB occurring at a rate $\Gamma$ is described by Eq.(\ref{meq}), with  $L=\sqrt{\Gamma}\sigma^-$. The evolution of a single fluorescent QB subject to  observation through direct photo-detection, involves  two projectors conditioned on no photo-detection $\tilde{\Omega}_0=e^{-\frac{\Gamma}{2}t}\ket{\uparrow}\langle \uparrow\vert+\ket{\downarrow}\langle \downarrow\vert$, and photo-detection $\tilde{\Omega}_1=\sqrt{1-e^{-\Gamma t}}\ket{\downarrow}\langle \uparrow\vert$, implying an asymptotic and instantaneous projection onto the ground state, respectively \cite{almeida}.
The projectors  $\{\tilde{\Omega}_k\}$  can be used to construct the Kraus operators for relaxation of a pair of  independent or interacting QBs (via $H_s$ in Eq.\ref{Hs}) \cite{carmichael,vedral2}. In either case, an initial state  $\ket{\psi_0}=\alpha\ket{\uparrow\uparrow}+\delta\ket{\downarrow\downarrow}$ will present no coherent dynamics. The description of relaxation subject to photon-detection is provided by  the Kraus operators $\Omega_{k=i+2j}=\tilde{\Omega}_i\otimes\tilde{\Omega}_j$ for independent QBs, while the projection after the first photon-detection into unexcited eigenstates $\ket{\psi_{\pm}}=\frac{1}{\sqrt{2}}(\ket{\uparrow\downarrow}\pm\ket{\downarrow\uparrow})$, in the interacting QBs case, is accounted  by the replacement $\Omega_{1,2}\rightarrow\Omega_{\pm}=\frac{1}{\sqrt{2}}(\Omega_1\pm\Omega_2)$. This latter decomposition assumes that the quantized vacuum mediated incoherent energy transfer is zero, which is exact whenever the induced dipole moments are orthogonal. Interestingly, the no-surveillance mixed state  $\rho(t)= \sum_{r}\Omega_{r}(t)\ket{\psi_0}\bra{\psi_0}\Omega^\dagger_r(t)=\sum_{s}\Omega_{s}(t)\ket{\psi_0}\bra{\psi_0}\Omega^\dagger_s(t)$, $r(s)=\{0,1(-),2(+),3\}$, highlights the ambiguous description of different quantum states which becomes invariant in an ensambled $\rho$ level. 

\begin{figure}
\includegraphics[width=2.4 cm]{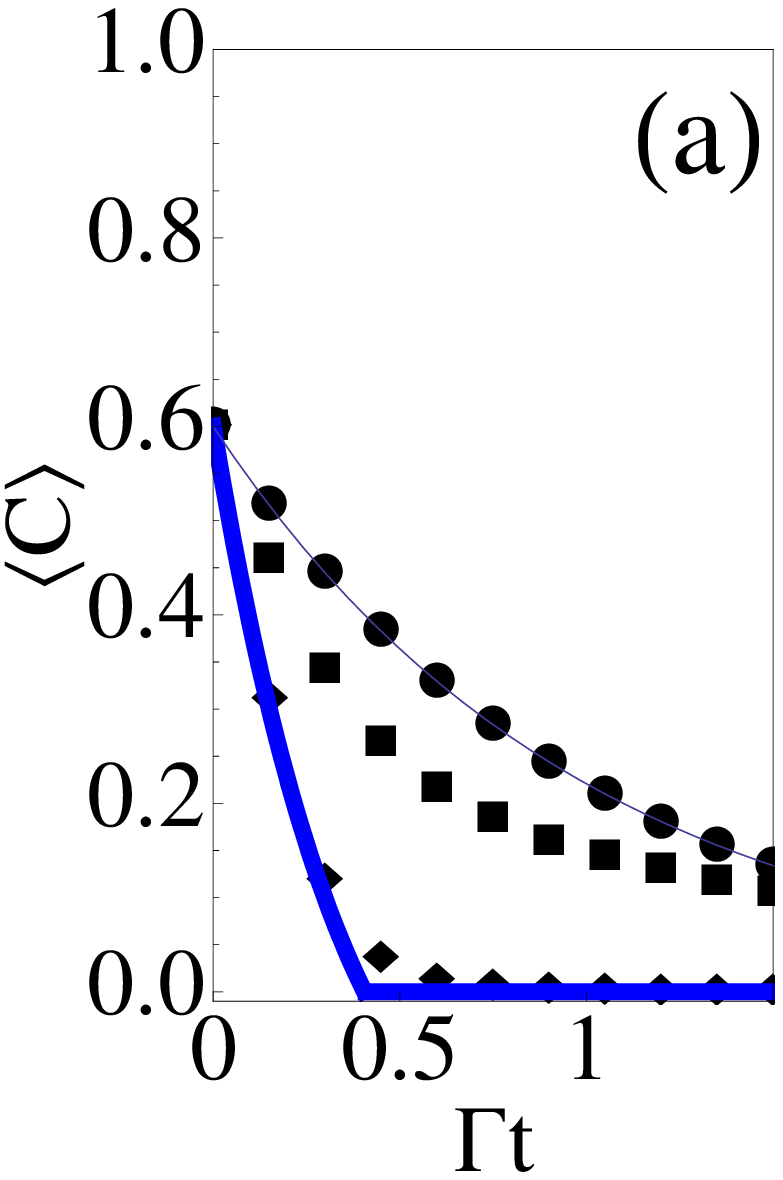}\hspace{-0 cm}
\includegraphics[width=5.6 cm]{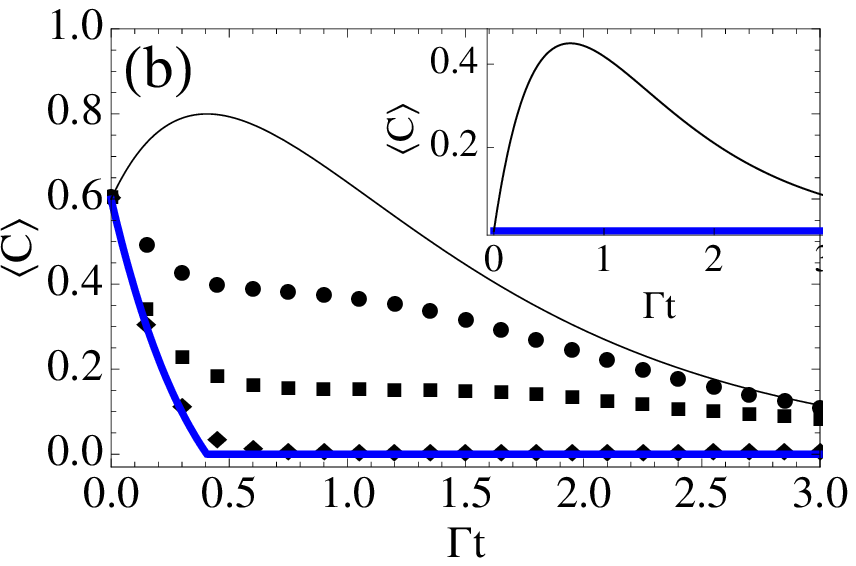}
\centering
\begin{minipage}{5 cm}
\includegraphics[width=4.6 cm,height=4.6 cm]{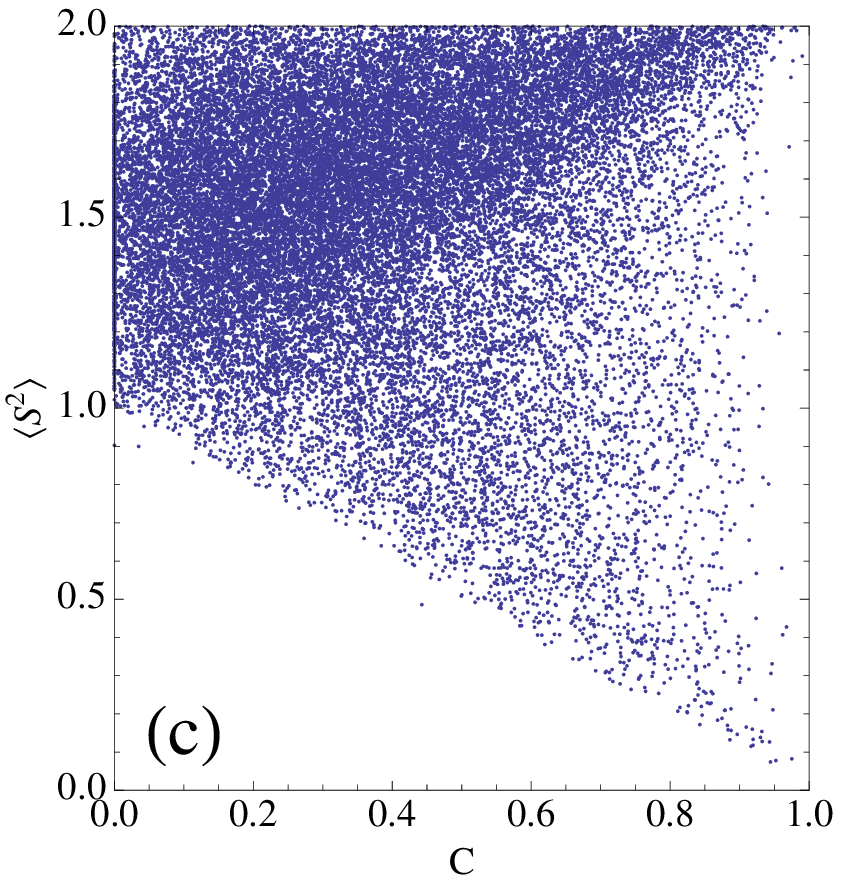}
\end{minipage}
\begin{minipage}{3.5 cm}
\includegraphics[width=3.5 cm]{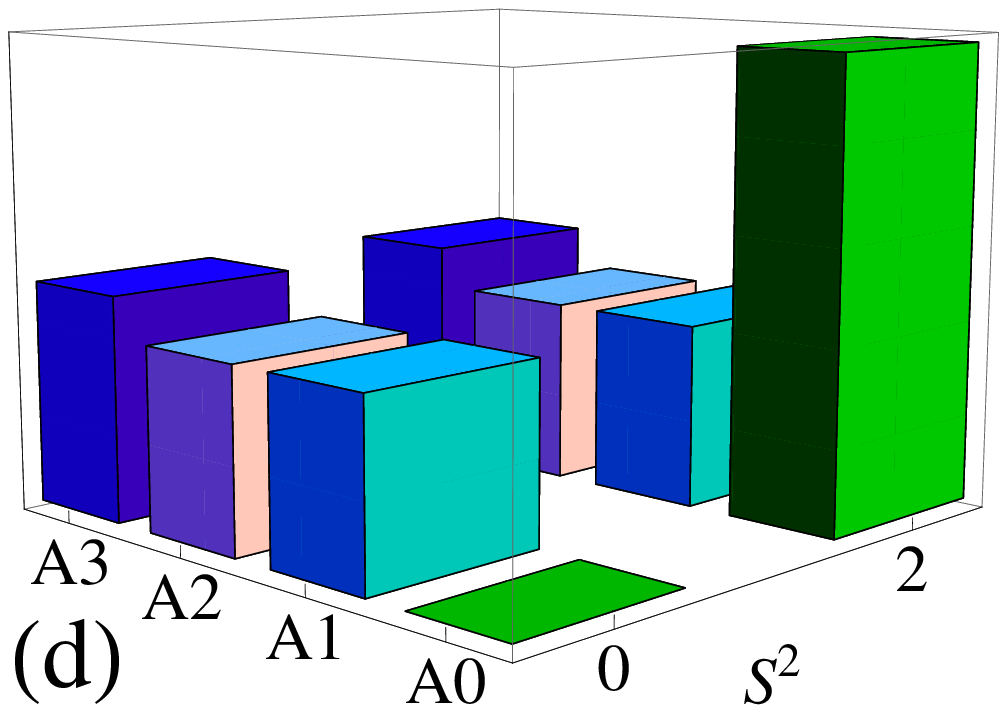}\\
\vspace{-0.1 cm}
\includegraphics[width=3.5 cm]{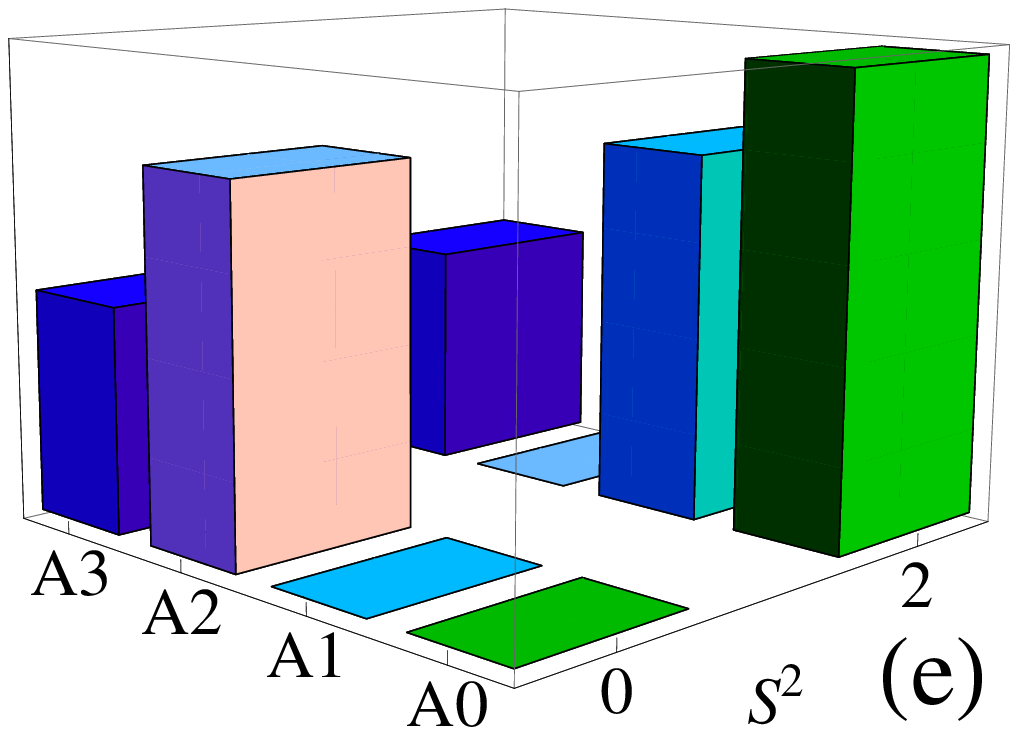}
\end{minipage}

\caption{(Color online)  Entanglement of independent QBs (a) and interacting QBs (b), initially prepared in the state $\ket{\psi_0}$, as quantified by the  analytical expression $\langle C\rangle$ based on Kraus decomposition  (thin, black), mixed state concurrence $C$  (thick, blue), and intermediate surveillance mean concurrence  $\langle C_n\rangle$  from numerical simulations with $n=\{2,5,50\}$ (circles, boxes and  diamonds, respectively). Inset in (b) shows  $\langle C\rangle_{\mbox{\footnotesize{int}}}$ and $C$ (equal to $\langle C\rangle_{\mbox{\footnotesize{ind}}}$) for $\alpha=1$. Witnessing of entanglement through  $S^2$ is possible as shown from the scatter plot (c), becoming an observable able to resolve the distinct entanglement features from the outcome  of $S^2$ on different channels for independent QBs (d) and interacting QBs (e) cases, at time $t=2 t_{\mbox{\tiny{ESD}}}$. $\alpha=\frac{3}{\sqrt{10}}$  such that $t_{\mbox{\footnotesize{max}}}=t_{\mbox{\tiny{ESD}}}$. }\label{sudden}
\end{figure}

{\it Beyond entanglement sudden death.-} The ESD  emerges when a pair of non-interacting QBs  subject to independent dissipative environments, are prepared in a superposition  $\ket{\psi_0}$ that  evolves into a mixture with concurrence $C(t)=\mbox{Max}[0, e^{-\Gamma t}\alpha^*(\alpha+e^{\Gamma t}(\delta-\alpha))]$, vanishing  in a finite time $\Gamma t_{\mbox{\tiny{ESD}}}=-\ln(1-\frac{\vert\alpha\vert}{\vert\delta\vert})$ whenever $\vert\alpha\vert\ge\vert\delta\vert$  \cite{eberly},   as shown in Fig.\ref{sudden}(a). As expected from the outlined invariance, the ESD equally occurs  with QBs interacting via Eq.(\ref{Hs})  (Fig.\ref{sudden}(b)). However, the use of Kraus operators when direct photo-detection is employed leads to the abrupt replacement of ESD by an  asymptotic decay \cite{carmichael} $\langle C\rangle_{\mbox{\footnotesize{ind}}}=2\vert\alpha\delta^*\vert e^{-\Gamma t}$ while  adding a non-monotonic term in the interacting case  $\langle C\rangle_{\mbox{\footnotesize{int}}}=2\vert \alpha\vert^2e^{-\Gamma t}(1-e^{-\Gamma t})+\langle C\rangle_{\mbox{\footnotesize{ind}}}$ to manifest the requirement  of visiting  the maximally entangled superpositions $\ket{\psi_\pm}$  that contrasts  when $\alpha=1$ to  the null result of $C$ or $\langle C\rangle_{\mbox{\footnotesize{ind}}}$ for all times (Inset Fig.\ref{sudden}(b)). The usual $C$ from a reduced density matrix approach is ill suited to describe this case by observing that $\langle C\rangle_{\mbox{\footnotesize{int}}}$  presents a maximum  at  $\Gamma t_{\mbox{\footnotesize{max}}}=\ln(2)-\ln(1+\frac{\delta}{\alpha})$ only  when ESD occurs, i.e. if $\vert\alpha\vert\ge\vert\delta\vert$.  Nevertheless, the comparison among  $C$ and $\langle C\rangle$, reflects the rather abrupt distinct situations of having none or all  information possible from the system, due to measurement over the environment.

 The effect of having different degrees of available information can be addressed if the averaging procedure is divided into subensembles of $n$ realizations, $\rho_n=\sum_{i=1}^{n}\frac{1}{n}\vert\psi_i\rangle\langle\psi_i\vert$, regarding the mean density operator in each subensemble as $n\rightarrow\infty$, while a gradual reduction of the subensemble size, until the best resolution is achieved when all subensembles have a single realization ($n=1$),  accounts on gaining progressively information concerning the times at which individual photons are detected \cite{PRA,thesisfcs}. Averaging the concurrence over all subensembles yields to $\langle C_n\rangle$. Figures \ref{sudden}(a)-(b) show how information loss due to the averaging of dephased trajectories, swiftly replaces  the asymptotic value to a finite time concurrence vanishing. However,  the robustness against information loss is  evident in the independent QBs case, which due to averages of factorable states absent of a fragile relative phase, present at $\langle C_2\rangle$ no noticeable difference with $\langle C\rangle$  (Fig.\ref{sudden}(a)), to be compared with the decrease of $\langle C_2\rangle$  against  $\langle C\rangle$ in  the interacting QBs (Fig.\ref{sudden}(b)) that highlight the delicacy of entanglement quantification due to averaging the dephased fully entangled states $\ket{\psi_\pm}$.  These results imply, ahead of conjectures for  entanglement beyond the ESD limit and its enhancement  by incoherent relaxation,  that  quantum correlations in $\rho$ are strongly vulnerable to ensemble averaging, and therefore the mixed state can give rise to an unphysical estimation of the entanglement dynamics in a single quantum system subject to noise.

In order to experimentally verify differences among these physical systems, we make use of  the photon time traces \cite{PRA}  by noting that ESD occurs irrespective of spectral overlap with independent QBs while transitions to $\ket{\psi_\pm}$ 
 differ by an energy of $g(r_{1,2})/2$ for interacting QBs. In this proposal the data from outcomes  of measuring  total angular momentum squared $S^2$, are binned according to the current photon time trace: channel 0 (A0) if no photon has been emitted, channel 1 and 2 (A1 and A2), are $S^2$ measures after the first spectrally resolved photon arrival but before channel 3 (A3), following a second photo-detection. The 
 observable $S^2$ may witness entanglement  since  its expected value $\langle S^2\rangle_\psi=2(\vert\alpha\vert^2+\vert\delta\vert^2)+\vert\nu+\gamma\vert^2$ for general pure states $\ket{\psi}=\alpha\ket{\uparrow\uparrow}+\gamma \ket{\uparrow\downarrow}+\nu \ket{\downarrow\uparrow}+\delta\ket{\downarrow\downarrow}$  ranges  between two and zero, while in fully mixed diagonal states $\rho=a\ket{\uparrow \uparrow}\bra{\uparrow \uparrow}+b\ket{\uparrow \downarrow}\bra{\uparrow \downarrow}+c\ket{\downarrow \uparrow}\bra{\downarrow \uparrow}+(1-a-b-c) \ket{\downarrow \downarrow}\bra{\downarrow \downarrow}$  the inequality $1\le\langle S^2\rangle\le2$ holds. Figure \ref{sudden}(c)   shows that  $\langle S^2\rangle<1$ only for entangled states, acting in this way as an EW,  from its calculation on density matrices constructed through random normalized complex numbers $\{\alpha,\gamma,\nu,\delta\}$,  with non-diagonal elements further multiplied by another complex uniformly distributed number within the unit circle such that  $\bra{i}\rho\ket{j}=\rho_{i,j}=\rho^*_{j,i}$ .  Hence, the possibility to witness entanglement relies on observing $\langle S^2\rangle_{A_k}<1$ in a given subensemble arising due to spectral resolution of detected photons.

Even though the mixed state average (not shown here) presents an asymptotical decay from $\langle S^2\rangle=2$ to $\langle S^2\rangle=1$ unable to verify quantum correlations, its single measurement read-out will yield to  the discrete eigenvalues $\{0,2\}$, the former able to witness entanglement. The binned discrete outcomes  of  $S^2$ for independent QBs (Fig.\ref{sudden}(d)), shows that no conclusion about entanglement can be made when $S^2$ is measured in A1 or A2, since in any of  these both channels $\langle S^2\rangle=1$, regarding the average from an equal amount of  quantized outputs 0 and 2,  expected from measurement of total angular momentum in a state $\ket{\uparrow\downarrow}$ or $\ket{\downarrow\uparrow}$.  On the other hand, Fig.\ref{sudden}(e) shows that this measurement is able to witness entanglement in the interacting QBs case, since 
all population in the $\ket{\psi_-}$ state provides $\langle S^2\rangle_{\mbox{\tiny{A2}}}=0$. Given that  these histograms  concern a time far beyond the ESD limit,  quantification of entanglement through $C$ should be disregarded as a loyal predictor of entanglement in general decompositions. Therefore, this case provides an example where witnessing quantum correlations is possible beyond the ensemble average when continuous environment observation provides the means to classify the outcomes from measuring a single quantum system.

{\it Conclusions.-} 
We have shown how the full statistical distribution of physically observable outcomes in single open quantum systems, can be used to predict/detect entanglement even when the simple ensemble averaged EWs, that rely on the dephasing-sensitive mixed state, yields to no entanglement at all. The usefulness of single read-out EW measurement record from individual quantum systems, paves the road to experimentally  witness entanglement beyond smeared out averages. We have also shown the need to shift attention from ensemble averages to physically meaningful decompositions by addressing explicitly the worked out examples of a two-spin system coupled to a classically diffusing third particle, on one hand, and a fluorescent system, on the other hand.

{\it Aknowledgements.-}  Proyectos Semilla of Facultad de Ciencias at Universidad de los Andes (2010-2011).

\end{document}